\documentclass{iucr}              

\usepackage{graphicx}
\usepackage{bm}
\usepackage{float}
\usepackage[usenames]{color}
\usepackage{epstopdf}
\graphicspath{{./figures/}}

     \paperprodcode{a000000}      
     \paperref{xx9999}            
     \papertype{FA}               

     \paperlang{english}          
     \journalcode{J}              
     \journalyr{2011}
     \journaliss{1}
     \journalvol{67}
     \journalfirstpage{000}
     \journallastpage{000}
     \journalreceived{0 XXXXXXX 0000}
     \journalaccepted{0 XXXXXXX 0000}
     \journalonline{0 XXXXXXX 0000}

\begin{document}                  



\title{On the estimation of statistical uncertainties on powder diffraction and small angle scattering data from 2-D x-ray detectors}
\shorttitle{Uncertainties on powder diffraction data from 2D detectors}


\author[a]{X.}{Yang}
\author[b]{P.}{Juh\'{a}s}
\cauthor[a,b]{S.~J.~L.}{Billinge}{sb2896@columbia.edu}{}

\aff[a]{{Department of Applied Physics and Applied Mathematics,
    Columbia University},
    \city{{New York}, New York, 10027, \country{USA}}}
\aff[b]{{Condensed Matter Physics and Materials Science Department,
    Brookhaven National Laboratory},
    \city{Upton, New York, 11973, \country{USA}}}









\maketitle                        

\begin{synopsis}
It is difficult to estimate statistical uncertainties from powder diffraction data collected on 2D detectors.  Here we compare different approaches and suggest best practices if the full variance-covariance matrix can be propagated and the best approximate approach when only the variances are propagated.
\end{synopsis}

\begin{abstract}
This paper explores optimal methods for obtaining one-dimensional (1D) powder pattern intensities from two-dimensional (2D) planar detectors with good estimates of their standard deviations. We describe methods to estimate uncertainties when the same image is measured in multiple frames as well as from a single frame. We show the importance of considering the correlation of diffraction points during the integration and the re-sampling process of data analysis. We find that correlations between adjacent pixels in the image can lead to seriously overestimated uncertainties if it is neglected in the integration process. Off-diagonal entries in the variance-covariance (VC) matrix are problematic as virtually all data processing and modeling programs cannot handle the full VC matrix.  We show the off-diagonal terms come mainly from the pixel splitting algorithm used as the default integration algorithm in many popular 2D integration programs, as well as from re-binning and re-sampling steps later in the processing. When the full VC matrix can be propagated during the data reduction, it is possible to get accurate refined parameters and their uncertainties at the cost of increasing computational complexity. However, as this is not normally possible, we explore the best approximate methods for data processing in order to estimate uncertainties on refined parameters with the greatest accuracy from just the diagonal variance terms in the VC matrix.
\end{abstract}


\section{Introduction}
\label{Introduction}

Two-dimensional (2D) X-ray detectors are now widely used with synchrotron and laboratory sources to obtain powder diffraction data~\cite{hebob;b;tdxrd09,livet;aca07}. Their use results in experimental throughputs several orders of magnitude faster than point or linear detectors~\cite{chupa;jac03}. Another advantage is that a 2D diffraction pattern contains much more information than the diffraction pattern obtained by conventional powder diffraction~\cite{hebob;b;tdxrd09}, though this information is rarely fully utilized, and the 2D image is reduced to 1D diffraction pattern by integrating around the Debye-Scherrer rings~\cite{dinne;b;pdtap08}. Regardless, accurately estimating the uncertainties on data is crucial, not least since they are used as weights in least-squares estimates of fitting parameters, and optimizing an unweighted least-squares will not, in principle, result in the maximum likelihood solution~\cite{bevin;dreaps92}. Knowing the uncertainties on the diffraction data is also the starting point for estimating the precision of parameters in the model~\cite{princ;b;mticams04,pecha;b;fopdascom05}.

Generally, there are two difficulties in estimating the errors on points in the 1D diffraction pattern collected from 2D detectors. First is how to estimate the variance of the raw counts in each pixel of the detector. Second is how to handle and propagate error correlations coming from the fact that intensities in neighboring bins in the 1D pattern may not be statistically independent. The degree of statistical correlation between points depends sensitively on choices made during the data reduction process as we describe below. Error correlations are often ignored but can have a significant effect on uncertainty estimates on refined parameters in a Rietveld or PDF refinement. 

Scattering is a quantum process and the counts of photons in a photon-counting detector follow a Poisson distribution which has a standard deviation of $\sigma_I \approx \sqrt{I}$~\cite{dinne;b;pdtap08}. However, most 2D detector technologies in wide use are integrating detectors, such as those based on CCDs or image plates, where the number of counts recorded in a pixel, after corrections for electronic noise and detector efficiencies, is proportional to, but not equal to, the number of detected photons and the uncertainties are therefore not simply the square root of the recorded counts. It is then not straightforward to estimate the uncertainties on raw intensities in a 2D image and they are often ignored.

If the detector gain is known, and it is assumed that non-Poissonian contributions to the noise (for example coming from shot noise in the electronics) are negligible, the uncertainties can be obtained by normalizing the intensities by the gain and taking the square root of the normalized counts as is done for photon counting detectors~\cite{hamme;hpr96,boldy;b;hpc10}. The challenge is to determine the detector gain in the particular experimental situation since it depends on X-ray energy and details of the readout.

A further complication comes from the fact that the 2D diffraction image is integrated, or averaged, around Debye-Scherrer rings to obtain the 1D diffraction pattern. If a pixel-splitting algorithm is used, this re-binning process may introduce statistical correlations between data in nearby points in the resulting 1D pattern. Accurate error estimations and error propagations should properly account for this, and in general it is necessary to propagate the full variance-covariance (VC) matrix, which quantifies not only the variance of the measured signal in each bin, but also the correlations between the errors on different points. Currently, most 2D integration software packages, such as Fit2D~\cite{hamme;esrf04}, Powder3d~\cite{hinri;zk06} and PyFAI~\cite{kieff;jpconfs13}, use the pixel-splitting algorithm as their default integration algorithm, which introduces statistical correlations between data in nearby bins in the 1D pattern even without subsequent rebinning or data resampling steps in the subsequent processing. As a result of all these issues, statistical uncertainties are rarely determined and propagated in powder data obtained from 2D detectors, which is a serious problem.

In this paper we assess different approaches for obtaining accurate uncertainty estimates on 1D diffraction patterns obtained from 2D data, including the degree of correlation in the errors quantified in the VC matrix. Different 2D to 1D reduction methods and different binning grids result in different levels of both uncertainties and error correlation between points in the 1D pattern. Currently, most data modeling programs, such as Rietveld refinement programs including GSAS~\cite{larso;unpub04}, FullProf~\cite{rodri;unpub90} and GSAS-II~\cite{toby;jac13} do not utilize the off-diagonal terms of the VC matrix, even when they are available. We describe protocols for obtaining accurate uncertainties and the full VC matrix on 1D powder patterns obtained from 2D diffraction data.  We have also implemented these protocols in python based software modules and an open-source program called {\sc SrXplanar} (https://github.com/diffpy/diffpy.srxplanar) This could be used in the integration step for obtaining more accurate estimated uncertainties on data that will be used for Rietveld refinement programs and for other applications that utilize 2D detectors for powder diffraction such as small angle scattering~\cite{barba;acie06} and PDF analysis~\cite{chupa;jac03}.  In practice, since most refinement programs cannot currently make use of the full VC matrix, we describe the best protocols for data processing to minimize statistical correlations in the 1D pattern.

\section{Theory}

\subsection{Overview of the process to integrate 2D images to 1D diffraction patterns}

The first step in the integration process is to calibrate the geometric parameters of the experiment such as incident beam center on the detector, sample-detector distance and tilt offsets of the detector. This is usually done by measuring the powder pattern from a calibration sample, such as National Institute of Science and Technology (NIST) standard silicon or ceria, where the structural parameters are known. Fitting routines for doing this are implemented in FIT2D~\cite{hamme;esrf04} and GSAS-II~\cite{toby;jac13}, for example.  We have used Fit2D to obtain these geometric parameters for our images.

The next step is conversion of a 2D image of pixels into a 1D histogram of bins. The bins are typically on a $2\theta$-grid, where $2\theta$ is the scattering angle, or they may be on a reciprocal space grid such as $Q = 4\pi\sin\theta/\lambda$ or $s = \sin\theta/\lambda$, where $\lambda$ is the X-ray wavelength and $\theta$ is the Bragg angle which is half the scattering angle. The image integration process consists of taking the intensities in the detector pixels and assigning them to the correct bins of the 1D array with the correct normalization.

We consider specifically data from isotropically scattering samples such as powder diffraction data and small angle scattering from untextured powders. In this case it is necessary to azimuthally average the counts around the conic sections of constant $2\theta$, where we note that the conic sections are usually circles but may be distorted to ellipses, due to detector tilts, but in a known way allowing the correct constant-$2\theta$ integration to be carried out. Except for excluded or masked pixels, each pixel or part thereof, is collected into the 1D bins according to the position of the pixel, and the intensity in the bin is calculated as an average or weighted average (depending on the specific algorithm used in reduction) of the intensities of pixels overlapping that bin.

One approach is to assign pixel counts proportionally to the coverage of the pixel in the bin~\cite{hamme;hpr96,hebob;b;tdxrd09}. This method assumes the intensity function is smoothly varying and estimates the counts that actually fall into the corresponding bin range according to
\begin{equation}
\label{eq;binsplitintensity}
O_{i} = \frac{{\sum_{j=1}^{N_i}} a_{ij} R_{j}}{\sum_{j=1}^{N_i} a_{ij}},
\end{equation}
where $R_{j}$ is the number of counts in the $j$th pixel and the sum is taken over all pixels overlapping the $i$th bin, $a_{ij}$ is the weight factor, which is usually proportional to the coverage of pixel $j$ to bin $i$, and $N_i$ is the number of pixels overlapping the $i$th bin. By assuming that the measurement of each pixel is independent, which is not always true as we discuss below, an estimate of uncertainty on the counts in the $i$th bin may be approximately estimated as the properly weighted standard deviation of those values,
\begin{equation}
\label{eq;binsplituncertainty}
\sigma_{i} = \sqrt{\left(\frac{\sum_{j=1}^{N_i} a_{ij} R_{j}^{2}}{\sum_{j=1}^{N_i} a_{ij}} - O_{i}^{2}\right) \times \frac{N_i}{N_i-1}},
\end{equation}
though this is not expected to give an especially accurate estimate and is fraught with problems.

There may also be intensity corrections that should be applied before the integration, for example to correct for polarization and geometrical effects as well as corrections for detector dark-current, flat-field, spatial distortion, and so on, in addition to removal of masked, dead or saturated pixels. Discussion of these issues is beyond the scope of this paper, though it is assumed here that they have been correctly handled.

\subsection{Statistical correlations between points in the 1D pattern}

Correlations between data points in the 1D pattern have several sources, including the correlation between adjacent pixels in the 2D detector, the algorithm used in re-binning the 2D image to a 1D sequence of intensities, and any re-sampling process that rebins intensities on to different grids during processing, for example, a $2\theta$ to $Q$ conversion or rebinning onto a grid suitable for fast Fourier transformation (FFT).

\subsubsection{Statistical correlation between adjacent pixels in the image}
\label{sec;pixelpixelcorrelation}

Depending on the design of the detector and the experimental conditions, intensities recorded in nearby pixels on 2D detectors may not be statistically independent. The origin of the correlation is complex and quite dependent on the detector design. Detail discussion on it is beyond the scope of this paper. Some detectors are designed to minimize cross-talk between pixels, such as pixel-array detectors and micromachined scintillators, but in the cases in this study, we found pixel-pixel correlations to be quite significant, though it is often ignored.

Different images taken with an identical experimental setup and recorded with the same incident flux will give statistically independent estimates of the scattering intensity. Thus, we can study the statistical distribution of uncorrelated data by making use of multiple frames. Uncertainty on the raw counts can be estimated in a single pixel by considering that same pixel in multiple frames and determining the standard deviation of the measured counts between all frames. Since the frames are statistically independent, this will give an accurate estimate of the standard deviation of the underlying distribution of counts in that pixel.  On the other hand, the intensity in the 1D pattern, obtained by integrating around the Debye-Scherrer rings, is influenced by the pixel-pixel correlation because neighboring pixels are often placed into the same 1D bin during the integration. In the results section we show that this effect is observed and may be large, indicating that pixel-to-pixel correlations were important in the case we studied, and should be taken into account in general.

If multiple frames are available, pixel-to-pixel correlations may be removed by making composite images by randomly selecting each pixel in the image from a different frame in the set of identical images.  We show that when the images are randomly sampled in this way the correct standard deviation is obtained on the 1D bins.

\subsubsection{Correlations due to the 2D to 1D integration algorithm}

Even if we assume the correlation between adjacent pixels can be ignored or is removed by sampling, the pixel-splitting method will introduce error correlations between neighboring bins, since each pixel contributes to more than one bin in the 1D pattern. Although it can be turned off in Fit2D~\cite{hamme;esrf04} and a non-pixel-splitting algorithm is used in GSAS-II~\cite{toby;jac13}, pixel-splitting algorithms are currently the default in many 2D integration software packages and users should be aware of this issue.

\subsubsection{Correlations due to re-sampling process during processing}

It is sometimes required to re-sample the 1D diffraction intensities onto a different grid, for example, from a $2\theta$-grid to a $Q$-grid. Similar to re-binning that takes place during the 2D integration, re-sampling introduces significant error correlations unless the re-sampling is from a fine to a coarse grid which minimizes bin sharing. However, this is undesirable in most cases since information in the data is lost when re-sampled to a coarser grid, making this a bad tradeoff in most circumstances.  We recommend a strategy that avoids rebinning by integrating the 2D image directly onto the final desired grid.

\subsection{Estimating uncertainties from 2D integrating detectors}

The first step in any error propagation process is to estimate the uncertainties on the raw data.  This is already difficult for integrating detectors such as image plates, CCDs and related detector technologies, since the uncertainties are not simply the square root of the counts as in a photon counting detector.  As we have discussed, the task is made more difficult due to error correlations between pixels in the image.

\subsubsection{Estimating uncertainties from multiple frames}
\label{sec;pixelSwitching}

We show below that directly calculating the standard deviation intensities after integration can give an overestimate due to pixel-pixel correlations. However, we can utilize the statistical independence of multiple frames to eliminate the effects of these correlations.  This is done by making compound images of statistically independent pixels by randomly exchanging pixels between images. If a proper rebuild algorithm is used, which does not duplicate or drop pixels during the rebuilding process, the final intensity will not change since the 1D patterns are averaged in the last step; however, the standard deviation that is estimated before the pixel shuffling will be different if nearby pixels are correlated.

In practice, the algorithm we use for the frame resampling is to pick two frames from the set at random and randomly select 50~\% of their pixels to switch. This process is repeated many times, 5000 in this particular case, until each pixel in composed frame is randomly chosen from all frames at the same position. We refer to this as the pixel-switching method.

\subsubsection{Estimating the uncertainty on the counts in each 2D pixel of a single frame}
\label{sec;esdSingleFrame}

When multiple frames are measured, the above approach works. However, it is experimentally expensive and requires extra care in keeping identical experimental conditions between exposures, or perhaps multiple frames are not available or not numerous enough. In general, it is desirable to have a method to estimate uncertainties on points in the powder pattern from a single frame.

For a 2D integrating detector the readout raw counts $R$ of one pixel (after corrections for dark current, flat field, and so on) is proportional to the number of X-ray photons $N$ that impinge on that pixel, with the constant of proportionality being the detector gain $G$,
\begin{equation}
\label{eq;countsFromN}
R = G N.
\end{equation}
If we know $G$, we can calculate the uncertainty on the raw counts by assuming the intensity has a Poissonian distribution,
\begin{equation}
\label{eq;errorFromN}
\sigma_R = G \sigma_N \approx G \sqrt{N}.
\end{equation}
However, the detector gain is usually hard to determine and not the same for different experiments. Here we would like to estimate it from a region of the detector with a uniform intensity, i.e. a region without sharp diffraction features. To determine the gain we consider a set of pixels in this region and assume that the underlying photon intensity is invariant over the region. The actual counts will therefore represent the statistical distribution function of the counts and the standard deviation may be determined.  The detector gain may then be determined by inverting Eq.~\ref{eq;errorFromN}.  In detail, the uncertainty of pixel $i$ in this region of uniform intensity is given by
\begin{equation}
\sigma_{R_i} = \sqrt{\frac{1}{N-1} \sum_{m=1}^{N} \left( R_m-\bar{R} \right) ^{2}},
\end{equation}
where $N$ is the number of pixels in the set of pixels in the vicinity of $i$th pixel, and the summation is taken over all the pixels in the neighborhood set. Knowing the raw counts and the standard deviation of a pixel, we can calculate the detector gain, by combining Eq.~\ref{eq;countsFromN} and Eq.~\ref{eq;errorFromN},
\begin{equation}
\label{eq;detectorgain}
G = \frac{\sigma_{R}^{2}}{R}.
\end{equation}
With the further assumption that, after proper flat-field corrections, the detector gain is the same for each pixel in the detector, we obtain the gain for all pixels by averaging the gain calculated from different uniform regions of the detector. Once we have the detector gain, the uncertainty of raw counts on all pixels can be estimated with Eq.~\ref{eq;errorFromN}.

In this process, an implicit assumption is that the non-Poissonian contributions to the noise are negligible. This is a reasonable assumption if the pixels have enough counts. It may appear that the pixel-pixel correlations will render this approach problematic since this method assumes uncorrelated intensities between pixels. However, we will show later that it works quite well and this is a reasonable approximation if the regions selected are relatively low in counts, away from regions with sharp diffraction features.

\subsection{Integration method}
\label{sec:integrationMethod}

Correlations of uncertainties in the 1D diffraction pattern depend on the integration method with the pixel-splitting algorithm introducing correlations between bins. To avoid it, we consider a non-pixel-splitting method for the re-binning process where the entire content of a pixel is assigned to a single bin based on the position of the pixel center. Because every pixel has the same weight, the counts in the bin are calculated as the average of all the pixels contributing to it.

This method leads to a less smooth line-profile on a fine grid. However, in most cases the effects are relatively small, and from the perspective of the uncertainties there is a large advantage that no error correlations are introduced by this process. We recommend this method be used for obtaining the most accurate uncertainties on 1D patterns and the least biased and best estimates of uncertainties on refined parameters.

\subsection{Propagating the full variance-covariance matrix}
\label{sec;propagateVCmatrix}

The most robust method for propagating the errors is to propagate the full VC matrix through the full data analysis chain. Here we present the mathematical approach, which is also implemented in our software program for image integration, {\sc SrXplanar}.

We use the common approach of treating all data reduction steps, such as integration or re-sampling steps, as linear operations~\cite{princ;b;mticams04} and express them in matrix form which is easier to generalize. Then, the full VC matrix is propagated using
\begin{equation}
\mathbf{Cov}_{\mathbf{o}} = \mathbf{T} \ \mathbf{Cov}_{\mathbf{c}} \ \mathbf{T}^{T},
\end{equation}
where $\mathbf{Cov}_{\mathbf{c}}$ is the VC matrix of the input data, $\mathbf{T}$ is the transformation matrix, and $\mathbf{Cov}_{\mathbf{o}}$ is the VC matrix of the output data. Full details and derivation of the expression can be found in Appendix~\ref{app;vcmatrixPropagation}

\subsection{Refinement with full VC matrix}

We would like to test the effect on the estimated uncertainties on refined parameters of ignoring off-diagonal terms in the VC matrix. Refinement using correlated data was previously studied by David~\cite{david;jac04}. However, we are not aware of a Rietveld refinement package that can handle the full VC matrix. We have written a refinement program that fits a single Bragg peak with a Gaussian function. The data are low resolution data from a 2D detector and the Gaussian lineshape works adequately, though not perfectly as we describe in Appendix~\ref{app;refinevcmatrix}.

We should also point out that the uncertainty we estimated is actually a measurement of precision, that is, a measure of the width of the confidence interval that results from random fluctuations in the measurement process~\cite{hahn;b;itc05}, rather than the more interesting accuracy, which is a measure of trust in the region of the underlying correct value. However, assessing the accuracy requires the knowledge of systematic errors and deficiencies in the model used in the refinement, which is beyond the scope of this paper.

\section{Experiment}
\label{ExperimentMethod}

Diffraction data from a standard ceria sample and from KFe$_2$As$_2$ were collected at the 11-ID-C beamline at the Advanced Photon Source (APS) at Argonne National Laboratory, using the rapid acquisition pair distribution function (RA-PDF) technique \cite{chupa;jac03} with beam size $0.5\times 0.5$~mm$^{2}$, temperature 100~K and wavelength $\lambda = 0.10798$~\AA.

A 2D Perkin Elmer amorphous silicon detector was used in the experiments. Dark images, with the X-ray shutter closed, of the same length of time were collected for each exposed frame and subtracted from the image. The corrected raw counts were recorded in a tiff format file and integrated with {\sc SrXplanar}. A correction for the solid angle subtended by each pixel is made during the integration, as well as for beam polarization. The pixel size of the detector was $0.2\times 0.2$~mm$^{2}$, and the distance between the detector and sample was 391.12~mm, which was obtained using FIT2D from a calibration sample in the usual way. This pixel size results in a $2\theta$ spread from $0.03 ^{\circ}$ (for pixels near the center of the detector) to $0.02 ^{\circ}$ (pixels at detector edge). We use $0.03 ^{\circ}$ as the approximate $2\theta$ size of one pixel in our discussion. It should be noted that FIT2D uses the $2\theta$ value of the pixel size as the default integration interval for the 1D function though this will vary with the particular integration software used.

\section{Results and discussion}

\subsection{Correlation between adjacent pixels}

In this subsection we explore the different ways described above for determining estimated standard uncertainties (e.s.u.'s), which is actually the estimated predicted uncertainties, on our 1D datapoints from the 2D images. In so doing, we discover the presence of significant statistical correlations between neighboring pixels in the images.

\begin{figure}
\caption{(a) Standard uncertainties estimated on 1D integrated pattern using EPP approach (a) and EBD approach (b). Curves are calculated using EPP50 and EBD50 (green), EPP50PS and EBD50PS (blue), and EPP1 and EBD1 (red), as described in the text below.}
\label{fig;stdmulone}
\includegraphics[width=1.0\textwidth, keepaspectratio]{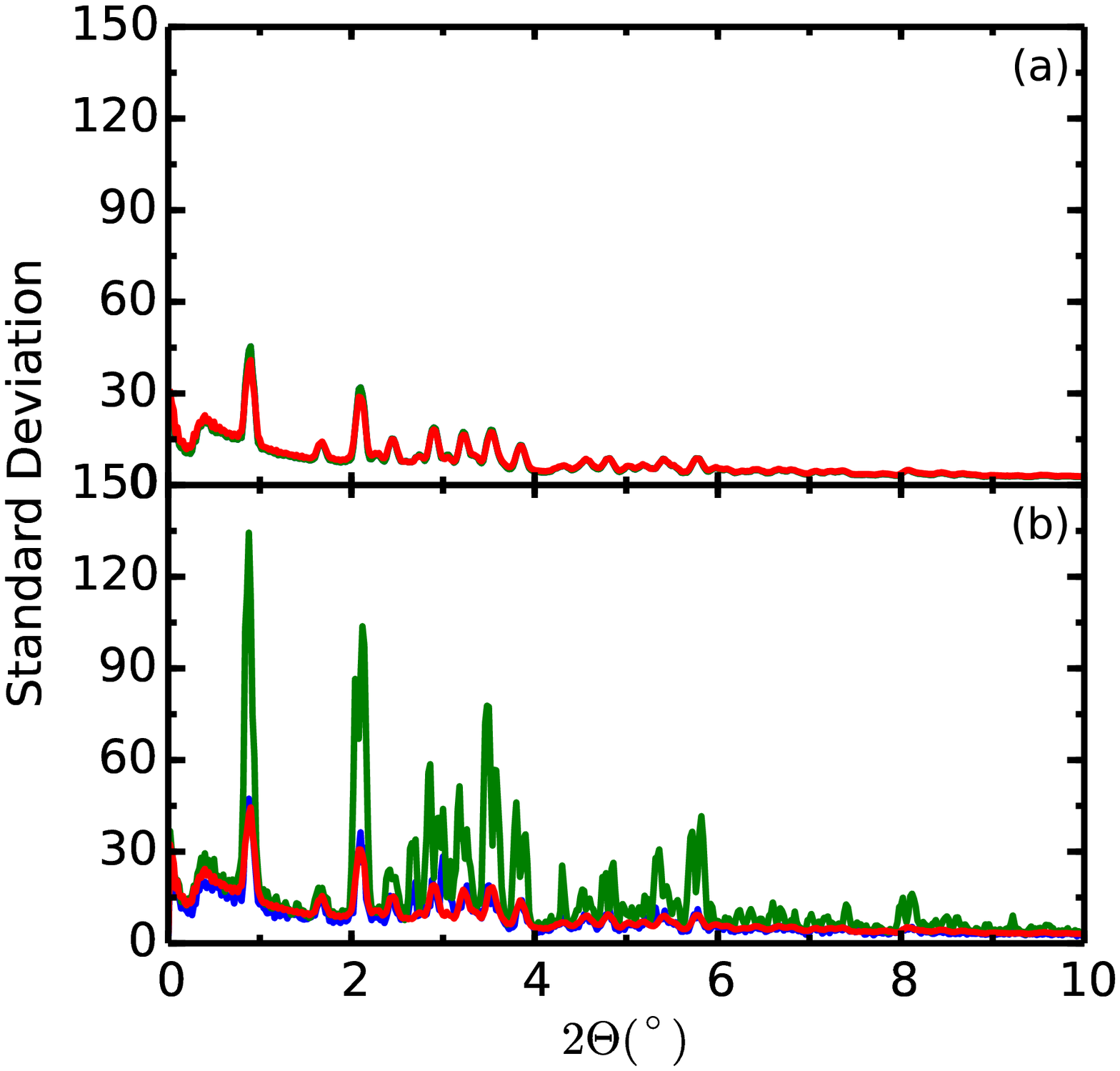}
\end{figure}

In our dataset we have 50 2D powder diffraction images, or frames, measured serially from the same sample with roughly constant incident flux. Each frame consists of $2048 \times 2048$ pixels in a square array. We can consider that each frame is an independent measurement of the 2D diffraction pattern of the sample. Each of the 2D images is then integrated azimuthally around the powder rings to obtain a 1D powder diffraction pattern of around 2000 intensity vs. $2\theta$ bins. Roughly speaking we can take two approaches to determine the e.s.u.'s: estimate the uncertainty on each pixel and propagate these uncertainties to the 1D bins (estimate pixel and propagate (EPP) approach), or alternatively, to integrate the pixel intensities into the 1D pattern first and then estimate the e.s.u of each bin directly (estimate bin directly (EBD) approach). In the EPP approach, the matrix transformation method described in Sec.~\ref{sec;propagateVCmatrix} and Appendix~\ref{app;vcmatrixPropagation} is used to propagate the uncertainties from the image to the 1D pattern. The e.s.u.'s on the bins in the 1D pattern are plotted in Fig.~\ref{fig;stdmulone}(a) using the EPP approach and in (b) using the EBD approach.

The green curves in Fig.~\ref{fig;stdmulone}(a) and (b) show the e.s.u.'s determined from the standard deviation of the observed intensities from the 50 independent measurements.  In Fig.~\ref{fig;stdmulone}(a) this is determined on a pixel by pixel basis then propagated to the 1D pattern (we call this the EPP50 method) and in Fig.~\ref{fig;stdmulone}(b) each image is integrated to a 1D pattern and the standard deviation is determined on each bin after the integration (we call this the EBD50 approach). These should give the same result as each other, but comparison of the green curves in Fig.~\ref{fig;stdmulone}(a) and (b) shows that they do not. Much larger uncertainties are estimated from the same data using the EBD50 approach than the EPP50 approach. One explanation for this behavior would be if the intensities in neighboring pixels in a single image were significantly statistically correlated. Taking the standard deviation before and after integration into 1D bins would then give a different standard deviation if these correlations were not taken into account when we propagate the uncertainty from 2D data array to 1D bins. To test this idea we incorporated a pixel-switching (PS) step (described in Section~\ref{sec;pixelSwitching}) before the data integration. This mixes the pixels between the images resulting in neighboring pixels in the image which must be independent. The resulting e.s.u's are shown as the blue curves in Fig.~\ref{fig;stdmulone}(a) EPP50PS, and (b) EBD50PS.  The blue and green curve in Fig.~\ref{fig;stdmulone}(a) must be the same as each other (the standard deviation is determined on precisely the same set of pixels) and they are. On the other hand, the EBD50PS approach (blue curve, Fig.~\ref{fig;stdmulone}(b)) now gives smaller e.s.u.'s that are in agreement with the EPP50 estimations.

To understand the slightly counterintuitive result that the EBD50 approach without pixel switching results in overestimated e.s.u's consider the following.  Assume a correlation does exist in each individual frame, such that when the measured counts in one pixel fluctuate up from the true expected value, there is a greater probability that the counts in its adjacent pixel also fluctuate up. If both pixels are placed into the same 1D bin, the counts in that bin are fluctuated even higher than if the pixels were uncorrelated.  A similar argument holds if the measured counts in the pixel fluctuate down, and the result is that the correlations amplify the fluctuations of the counts in the 1D bin.

This result shows that, in our case, significant statistical correlations exist between neighboring pixels in the images, i.e., for the Perkin-Elmer amorphous silicon detector we used, this effect is significant.  It also shows that when estimating e.s.u.'s from the standard deviation of intensities from multiple images, an EPP approach should be used.

We now test whether accurate e.s.u.'s can be obtained using data from just a single image by using the single-frame method described in Section~\ref{sec;esdSingleFrame} (we call these EPP1 and EBD1, though in this case EPP1 and EBD1 actually amount to being same procedure). The Red curve in Fig.~\ref{fig;stdmulone}(a) and (b) shows the e.s.u's when the non-splitting integration is used. They are in good agreement with those obtained by the EPP50, EPP50PS, and EBD50PS method, which suggests that, at least for this detector, the single-frame method for estimating the e.s.u.'s is sufficiently accurate, as well as being much more convenient.

Given that we have shown that neighboring pixels are statistically correlated in our data, it is somewhat surprising that the single-frame estimate does work, since the detector gain is determined from the standard deviation of pixels in the same neighborhood in the image. The possible reason is that the pixel-pixel correlation is weak in the low intensity region without sharp diffraction features, so the detector gain estimated from the distribution of intensities in that region is correct. This is partially supported by the fact that both the EPP50 and EBD50 estimates are similar in regions of the pattern with low intensity diffuse scattering, which is precisely the regions used in the detector gain estimation. The single-frame method also relies on the assumption that, after the flat-field correction, the detector gain is the same for all pixels. To test this assumption, we calculated the detector gain of each pixel using Eq.~\ref{eq;detectorgain}, where $\sigma_{R}^{2}$ is calculated using intensities on the same pixel from 50 frames. We found that in the case we studied, the assumption of equal gain holds very well. The good agreement between the EPP50 and EPP1 estimates implies that the single-frame method produces accurate uncertainty estimates. To further verify this, we estimated uncertainties from several different single-frames in the set and compared them to each other. The results are very similar (not shown in the figure), further validating this approach.

\subsection{Splitting vs non-splitting integration method}
\label{sec;sp_vs_nonsp_method}

\subsubsection{Integrated intensity profile}

\begin{figure}
\caption{1D diffraction intensity integrated with: (a) $\Delta 2\theta = 0.002^\circ$ and (b) $\Delta 2\theta = 0.02^\circ$. We integrated the 2D diffraction image with the pixel-splitting (green) and non-pixel-splitting (blue) methods, respectively. The $2\theta$ value of the detector pixel size is equal to $0.03^\circ$ in this case.}
\label{fig;smoothvsjagged}
\includegraphics[width=1.0\textwidth, keepaspectratio]{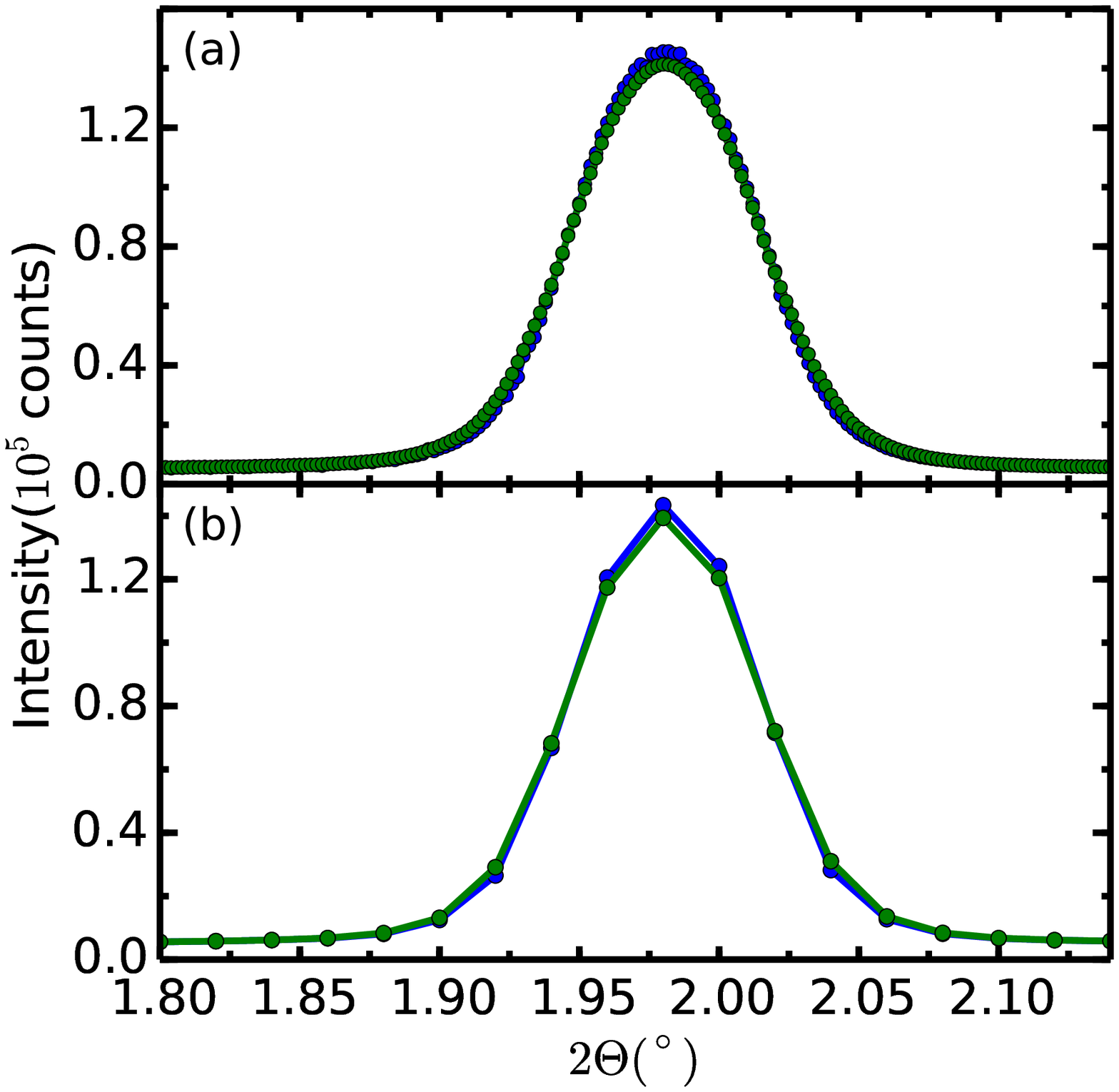}
\end{figure}

In Fig.~\ref{fig;smoothvsjagged} we compare the diffraction patterns obtained from the same image when the integration is carried out using the pixel-splitting and non-splitting method at two representative intervals, $\Delta 2\theta = 0.002^\circ$ and $\Delta 2\theta = 0.02^\circ$.
The non-splitting method data have the same integrated peak intensity but appear less smooth, though slightly narrower. In fact, the pixel-splitting method is a {\it de facto} smoothing method, which gives smoother data at the expense of resolution. The basic assumption is that the diffraction intensity is uniformly distributed within each pixel, which makes the intensity change between bins smoothly due to the pixel sharing at bin edges. It also makes some pixels at bin edges in the peak center bin give part of their counts to the shoulder, which increases the intensity of shoulder and decreases the intensity of center resulting in a slight peak broadening.
In comparison, the non-splitting method assumes the entire intensity belongs to one bin even if part of that pixel falls into other bins. When data are binned on a very fine grid where the bin size is much smaller than the pixel size, the non-splitting method is under-sampled and some bins will be empty. However, there is little advantage to binning the data as finely as this since the resolution is limited by the pixel size at the very least, and in a well designed experiment the intrinsic resolution will be worse than the pixel width. When a large bin width is used, the number of pixels that cross the bin edge is small compared to the total number of pixels fully in the bin, and therefore the difference between two methods is small, though still significant when the bin width matches the pixel width, as evident in Fig.~\ref{fig;smoothvsjagged}.

Of course, the non-splitting method does not introduce any statistical correlations and is preferred for that reason.

\subsubsection{Propagating the full variance-covariance matrix}~

\begin{figure}
\caption{VC matrix of the diffraction intensity integrated with: (a) non-splitting method and $\Delta 2\theta = 0.03^\circ$, (b) splitting method and $\Delta 2\theta = 0.03^\circ$, (c) splitting method and $\Delta 2\theta = 0.006^\circ$, and (d) splitting method and $\Delta 2\theta = 0.08^\circ$. The $2\theta$ value of the pixel size is equal to $0.03^\circ$ in this case. Only the non-splitting method gives a diagonal VC matrix.}
\label{fig;vcmatrix}
\includegraphics[width=1.0\textwidth, keepaspectratio]{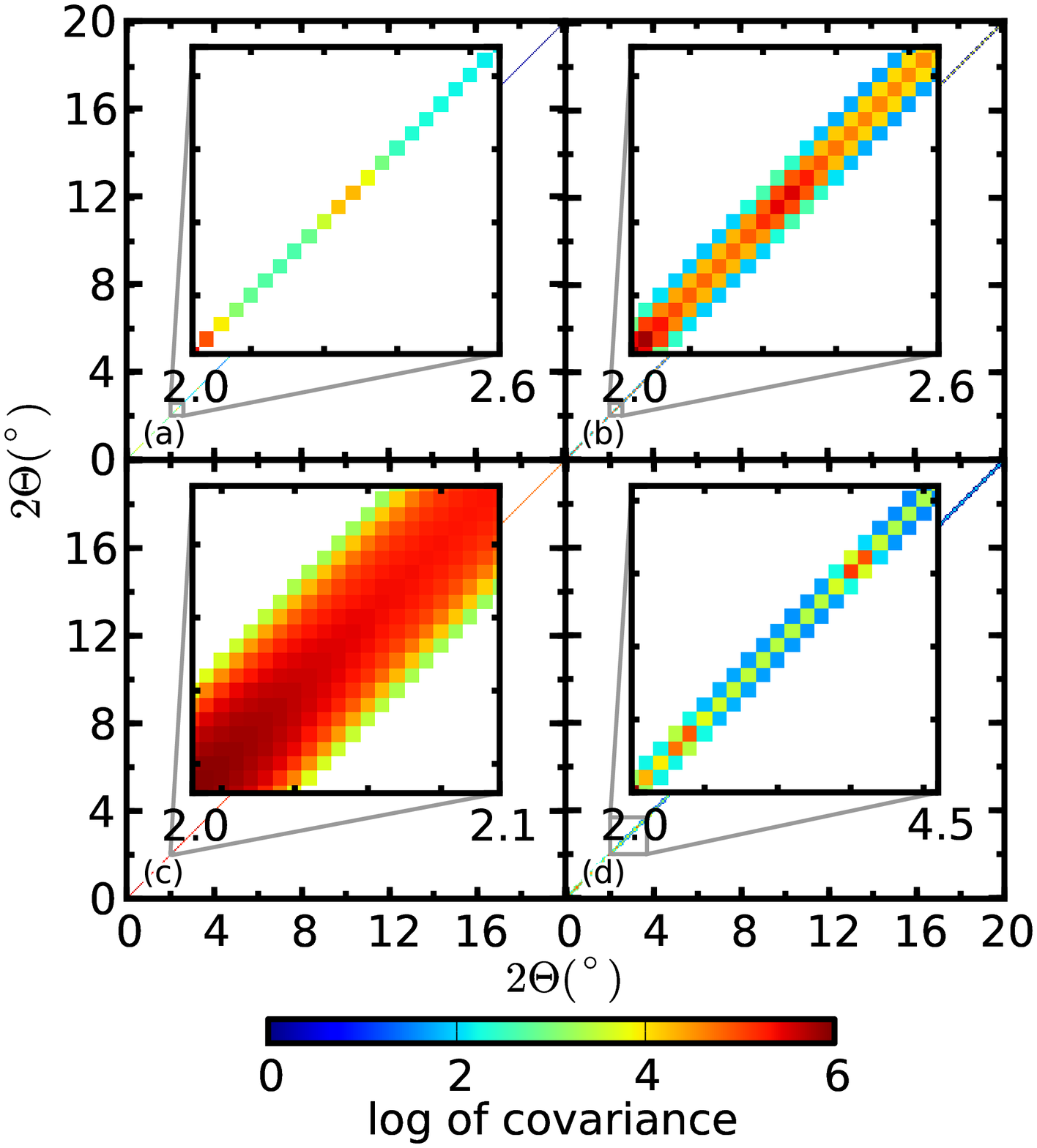}
\end{figure}

In Fig.~\ref{fig;vcmatrix} we show a false-color image of the VC matrix with the integration done in different ways.

As expected, the VC matrix obtained by the non-splitting integration method is diagonal, while that obtained by the splitting method is mostly diagonal but with a ridge along the diagonal. The width of the ridge depends on the bin width compared to the pixel size, with a broader ridge from the smaller bin-size, and a nearly diagonal ridge when the bin size is approximately equal to pixel size.

\subsection{Influence on the VC matrix of a re-sampling of the 1D pattern}
\label{sec;VCmatrix_resampling}

Here we study how re-sampling the 1D function onto a new grid affects the VC matrix. We tested two types of re-sampling steps which are common in data analysis. One is to re-sample the diffraction intensity from a regular $2\theta$-grid to a regular $Q$-space grid. The other is to re-sample the data on to another regular grid in the same integration space.

\begin{figure}
\caption{Selected range of VC matrix of diffraction intensity directly integrated or re-sampled to corresponding integration grid. The original intensity was integrated to a regular $2\theta$-grid with $\Delta 2\theta = 0.03 ^\circ$. The integration grids are: (first column, a, d, g, j) regular $Q$-space grid with same length of original intensity; (second column, b, e, h, k) regular $2\theta$-grid with $\Delta 2\theta = 0.01 ^\circ$; (third column, c, f, i, l) regular $2\theta$-grid with $\Delta 2\theta = 0.05 ^\circ$. The 2D diffraction pattern was (first row, a, b, c) directly integrated with non-splitting method, (second row, d, e, f) re-sampled from data integrated with non-splitting method, (third row, g, h, i) directly integrated with splitting method, (fourth row, j, k, l) re-sampled from data integrated with splitting method.}
\label{fig;resample}
\includegraphics[width=1.0\textwidth, keepaspectratio]{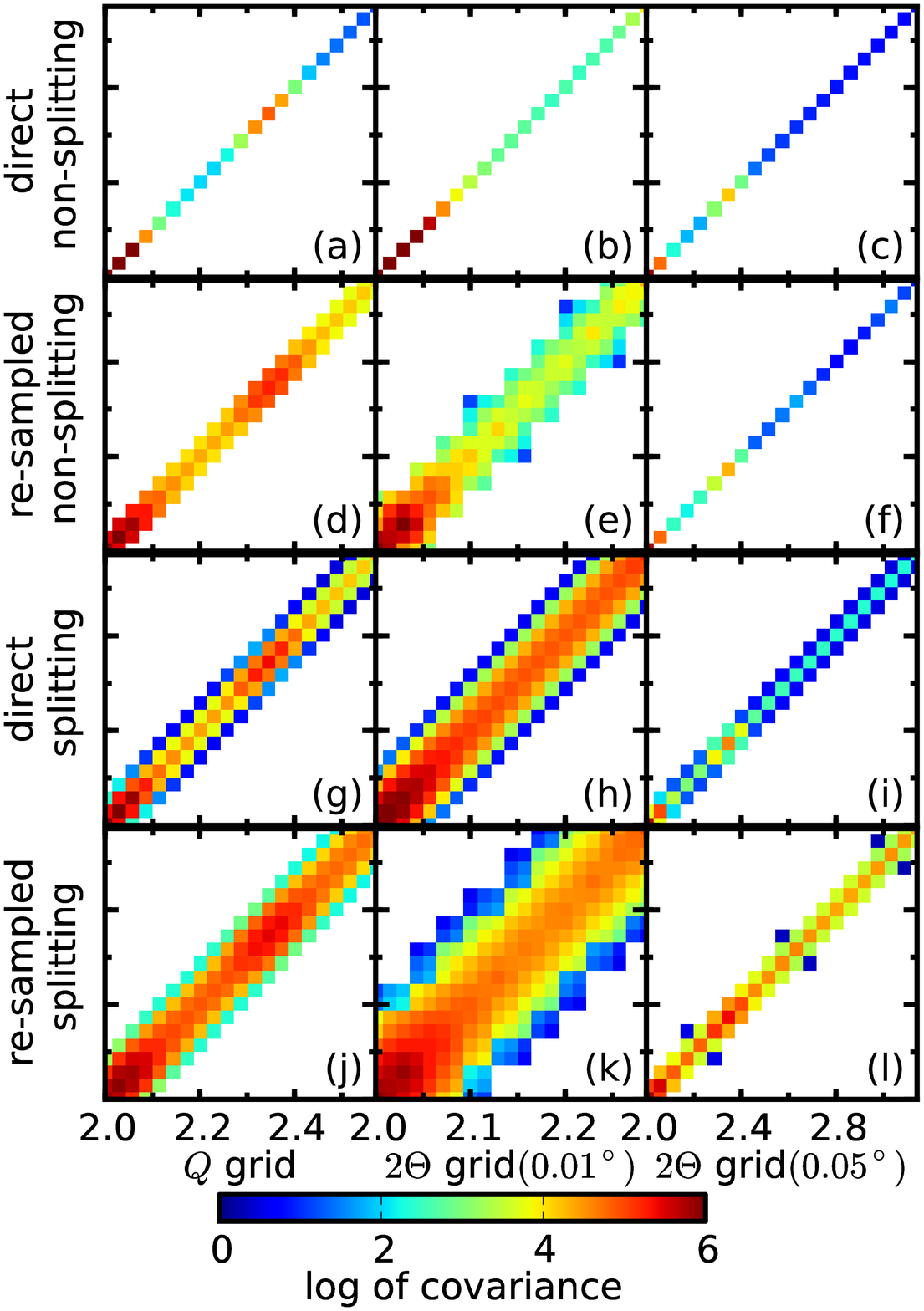}
\end{figure}

We started with the diffraction intensity initially integrated in $2\theta$-space with $\Delta 2\theta = 0.03^\circ$ and using the non-splitting method and the splitting method. The data were then re-sampled on to a $Q$-space grid of the same length, and alternatively onto another $2\theta$-space but with interval equal to $0.01^\circ$ and $0.05^\circ$. As a comparison, we have also integrated the 2D image directly onto the same $Q$-space grid and $2\theta$-grids. The results are shown in Fig.~\ref{fig;resample}.

Compared to data that were directly integrated, data that were re-sampled from a $2\theta$-grid to a $Q$-space grid, or to a finer grid, show increased statistical correlations, though we note that interpolation to a coarser grid does not induce significant additional correlations in the resulting data.

In summary, direct integration of the 2D data onto the final grid results in the smallest correlations and should be the best practice. Usually there is no need to re-sample if one carefully chooses the integration grid during 2D integration process. This result is not a surprise, but it is often not followed in practice and worth mentioning explicitly. It is especially relevant in measurements like small angle scattering or PDF where the subsequent data processing is conducted on a $Q$-grid. For Rietveld refinement, McCusker suggested the optimal binning for Rietveld is where there are 6-10 points over the FWHM of a peak~\cite{mccus;jac99}. If during processing it is determined that a new grid is needed, it is preferable that the software can reintegrate the original images onto the new grid.

\subsection{Structural refinements utilizing the full VC matrix}

Most powder diffraction data are modeled by some kind of peak fitting process and we would like to know the effects on refined parameters of neglecting the error correlations under discussion. Since no Rietveld programs available to us could handle the full VC matrix, we wrote our own simple fitting program and fit the second peak in a test data-set from a ceria sample ($1.80^{\circ} < 2 \theta < 2.15^{\circ}$) to a Gaussian shape function given by $y = \frac{A}{\sigma \sqrt{2 \pi}} \exp(-\frac{(x-x_0)^2}{2 \sigma ^2}) + B$, where $A$ and $x_0$ are the integrated intensity and peak position, respectively, and $B$ is a constant background. In the refinement, we refine $A$, $x_0$, $\sigma$ and $B$.

Here we discuss the effect on the values and estimated uncertainties of refined parameters due to different data reduction choices: (a) fitting data integrated with the non-splitting method (note that the VC matrix obtained by the non-splitting method is already diagonal), (b) fitting data integrated with the splitting-method but considering the full VC matrix in the fit, and (c) fitting data integrated with the splitting-method but only using the diagonal of the VC matrix (this approach is strictly incorrect but is the {\it de facto} current approach in most studies). We also extended the study to explore the effect on the values and estimated uncertainties of using a unit weight matrix in the least-squares equation instead of using the propagated statistical uncertainties. To this end, we explored (d) fitting data integrated with the non-splitting method but using a unity weight matrix and (e) as (d) except the fits were on data integrated with the splitting method. The uncertainties are estimated using Eq.~\ref{eq;covWithVC} and Eq.~\ref{eq;covWithoutVC}. In summary, we want to know the effect on the estimated uncertainties of ignoring off-diagonal covariance terms in the analysis (i.e., (c)) and the effect on the refined values themselves of using the wrong weights in the minimization (i.e., (d) and (e), and to a lesser extent (c))

The results of the fitting are shown in Table~\ref{tab;vcrefine}. We refine data that were processed with two integration intervals, an interval much smaller than the pixel size ($\Delta 2\theta = 0.13 \times \Delta 2\theta_0 = 0.004^{\circ}$) and an interval equal to the pixel size ($\Delta 2\theta = 1.0 \times \Delta 2\theta_0 = 0.03^{\circ}$).  In the latter case the data correlations are much smaller as discussed earlier.
\begin{table}
\caption{Peak fit results. Refinements used (a) data integrated with the non-splitting method, (b) data integrated with the splitting method and considering the full VC matrix, (c) data integrated with the splitting method but ignoring the off-diagonal elements of the VC matrix, (d) data integrated with the non-splitting method and refined with a unity VC matrix, and (e) data integrated with the splitting method and refined with a unity VC matrix. $\Delta_1$ and $\Delta_2$ are uncertainties on the refined parameter estimated using Eq.~\ref{eq;covWithVC} and Eq.~\ref{eq;covWithoutVC}, respectively. Please refer to the main text for their meaning. For (d) and (e), $\Delta_1$ is not available since unity weight matrix was used in the refinement.}
\label{tab;vcrefine}
\centering
\begin{tabular}{lrrrrr}

\multicolumn{6}{l}{$\Delta 2\theta = 0.013 \times \Delta 2\theta_0 = 0.004^{\circ}$} \\
    & (a) & (b) & (c) & (d) & (e) \\
\hline
\multicolumn{6}{l}{Peak position ($^{\circ}$)} \\
Value          & 1.980729 & 1.980727 & 1.980725 & 1.980770 & 1.980754 \\
$\Delta_1$     & 0.000004 & 0.000004 & 0.000002 & -        & -        \\
$\Delta_2$     & 0.000109 & 0.000084 & 0.000078 & 0.000124 & 0.000087 \\
\hline
\multicolumn{6}{l}{Peak width ($^{\circ}$)} \\
Value          & 0.030897 & 0.032103 & 0.032095 & 0.031032 & 0.032148 \\
$\Delta_1$     & 0.000003 & 0.000003 & 0.000002 & -        & -        \\
$\Delta_2$     & 0.000125 & 0.000097 & 0.000091 & 0.000124 & 0.000095 \\
\hline
\multicolumn{6}{l}{Peak intensity} \\
Value          & 10977.4  & 10947.2  & 10994.1  & 11017.5  &  11012.6 \\
$\Delta_1$     & 1.5      & 1.4      & 0.5      & -        & -        \\
$\Delta_2$     & 47.0     & 35.0     & 33.0     & 46.0     & 32.0     \\
\hline
\multicolumn{6}{l}{Durbin-Watson $d$-statistic ($Q$=1.4)} \\
Value          & 0.170    & 0.092    & 0.097    & 0.291    &  0.189   \\
\hline
\hline
 & & & & & \\
\multicolumn{6}{l}{$\Delta 2\theta = 1.0 \times \Delta 2\theta_0 = 0.03^{\circ}$} \\
    & (a) & (b) & (c) & (d) & (e) \\
\hline
\multicolumn{6}{l}{Peak position ($^{\circ}$)} \\
Value          & 1.980709 & 1.980706 & 1.980700 & 1.980722 & 1.980731 \\
$\Delta_1$     & 0.000004 & 0.000004 & 0.000004 & -        & -        \\
$\Delta_2$     & 0.000203 & 0.000167 & 0.000167 & 0.000202 & 0.000166 \\
\hline
\multicolumn{6}{l}{Peak width ($^{\circ}$)} \\
Value          & 0.031626 & 0.032784 & 0.032873 & 0.031637 & 0.032841 \\
$\Delta_1$     & 0.000003 & 0.000003 & 0.000003 & -        & -        \\
$\Delta_2$     & 0.000234 & 0.000195 & 0.000195 & 0.000233 & 0.000233 \\
\hline
\multicolumn{6}{l}{Peak intensity} \\
Value          & 10932.2  & 10991.8  & 11003.0  & 10943.2  &  11006.4 \\
$\Delta_1$     & 1.5      & 1.5      & 1.2      & -        & -        \\
$\Delta_2$     & 85.0     & 70.0     & 70.0     & 126.0    & 99.0     \\
\hline
\multicolumn{6}{l}{Durbin-Watson $d$-statistic ($Q$=1.4)} \\
Value          & 1.714    & 1.553    & 1.800    & 1.515    &  1.527   \\
\end{tabular}
\end{table}
\begin{figure}
\caption{Integrated intensity (blue), fitting results (red), and difference curve (green) of second peak integrated with $\Delta 2\theta = 0.004 ^\circ$. Refinement is performed on (a) data integrated with the non-splitting method, (b) data integrated with the splitting method and considering the full VC matrix. Part of the integrated intensity and the difference curve are zoomed 10x for better visualization}
\label{fig;fitting}
\includegraphics[width=1.0\textwidth, keepaspectratio]{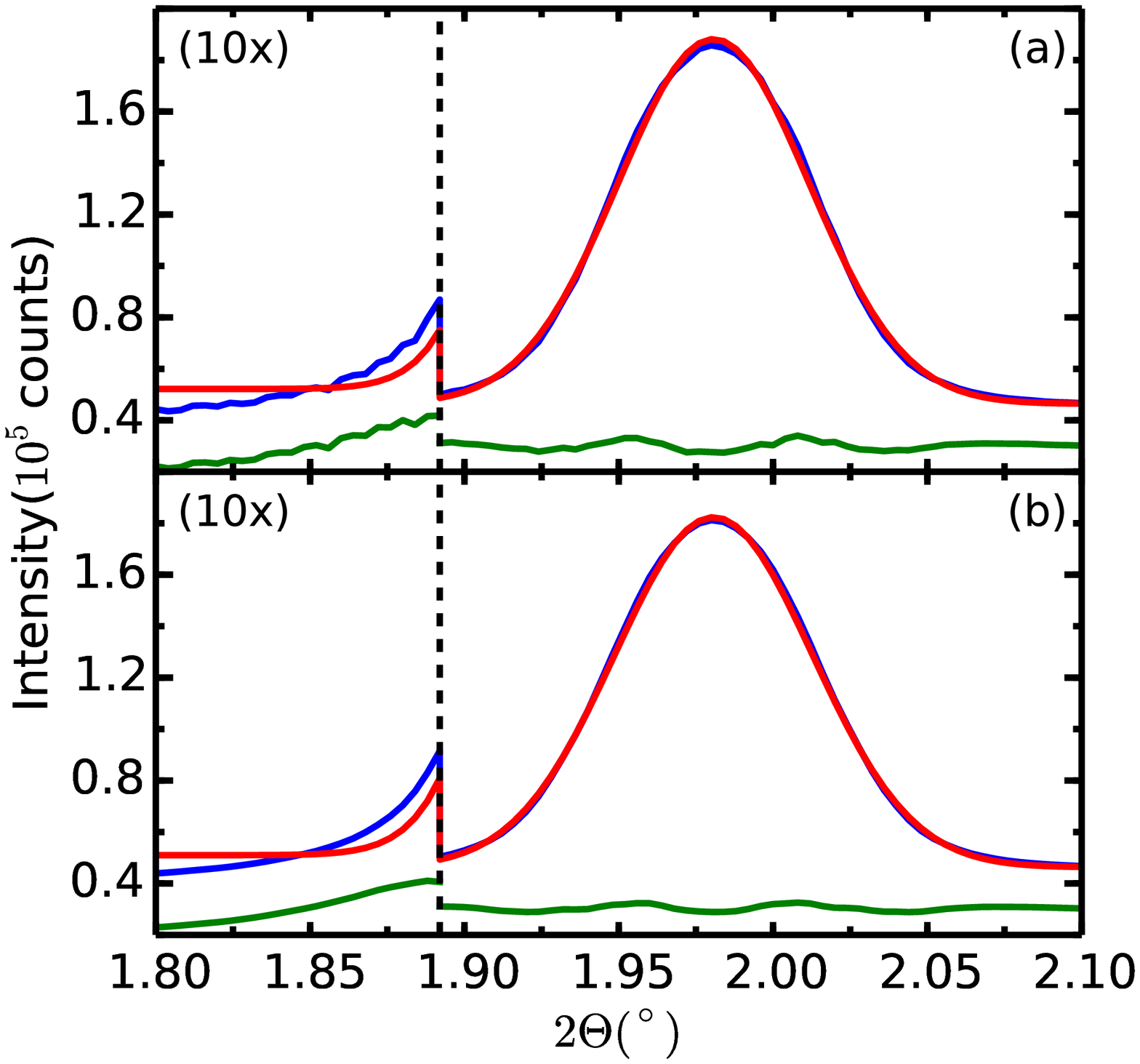}
\end{figure}

The uncertainties of refined parameters are calculated using two different metrics. $\Delta_1$, calculated using Eq.~\ref{eq;covWithVC}, only counts the contribution to the uncertainty due to statistical errors. On the contrary, $\Delta_2$, calculated using Eq.~\ref{eq;covWithoutVC}, is a less statistically justifiable metric but is widely used in refinement programs to estimate uncertainties. To some extent it takes into account model errors since it scales the estimated uncertainties by the residual between the experimental and calculated data.

For all refined parameters, $\Delta_1 $ is much smaller than $\Delta_2$ by one order of magnitude. This indicates, in our specific case, that model errors dominate the residual, i.e. inconsistencies between the calculated and data patterns due to deficiencies in the model are contributing much more to the residuals than the statistical errors.  This is a common situation in many Rietveld refinements.  However, the systematic error does not have the same effect on all the different refined parameters. For the peak position, when the statistical errors are correctly handled, i.e. method (a) and (b), or partially correctly handled, i.e. method (c), the refined results are very close to each other and all lie within $\Delta_1$ of each other. On the other hand, ignoring the statistical uncertainty in the refinement, i.e. method (d) and (e), leads to a significant deviation compared to (a-c). The deviation is much larger than $\Delta_1$ but smaller than $\Delta_2$. It implies that, in this case, the uncertainty of the peak position is dominated by the statistical uncertainties and the $\Delta_2$ uncertainty estimate grossly over-estimates the true uncertainty. This is because the peak in the data, whilst not pure Gaussian, is symmetric, and so fitting it with a symmetric function such as a Gaussian does not bias the refined value. This is supported by looking at the difference curve in Fig.~\ref{fig;fitting} which has a rather symmetric `w' shape indicating that the misfit is symmetric. On the other hand, the $\Delta_2$ estimation of the uncertainty does a much better job than the purely statistical $\Delta_1$ for the other refined parameters in the Gaussian.  Thus, cases where the fit residual is dominated by model deficiencies, $\Delta_2$, are still a better measure of uncertainty than $\Delta_1$ although they cannot be relied on to give accurate uncertainty estimates on refined parameters.

In principle, the statistical errors have been correctly handled using both methods (a) and (b), which should therefore be in good agreement with each other for each bin-size.  Indeed, the statistical uncertainties are in good agreement using the two methods. The agreement of (a) and (b) also implies that the splitting and non-splitting method carry the same amount of information when the full VC matrix is used in the analysis. On the other hand, when the off-diagonal terms in the VC matrix are ignored (the case (c)) the statistical uncertainties on refined parameters are consistently underestimated.

For the peak width, the situation is more complex, since the splitting method actually broadens the peaks and so it is not possible to compare the results of the splitting and non-splitting methods with each other, as discussed in Sec.~\ref{sec:integrationMethod}. This is clearly seen in Table~\ref{tab;vcrefine} where the splitting method results (b), (c) and (e), are all broader than the non-splitting method (a) and (d). The broadening is considerably larger even than the large $\Delta_2$ uncertainty estimates.  This indicates that the smoothing effects of using the splitting method are much larger than the magnitude of systematic error caused by peak profile mismatch and should be therefore considered.

For the case of the peak intensity, we do not expect to see effects of the pixel-splitting protocol used since this preserves the integrated intensity.  However, we see that the mismatch of the model peak profile to the data introduces large uncertainties on the peak intensity. The absolute difference between each value is $\sim 10\times$ larger than $\Delta_1$ though $\Delta_2$ seems to do a better job of estimating these uncertainties. Since the refined values were from the same original data-set, and it is not completely clear {\it a priori} which method will give the best estimates, the difference between the results from the different integration methods gives a measure of our actual uncertainty on the values of the refined parameters. In the case of the peak intensity, the  errors coming from the inadequate model dominate the real uncertainties.

The serial correlation~\cite{hill1;jac87,andre;jac94,Berar;jac91} in each refinement was characterized using the Durbin-Watson $d$-statistic~\cite{durbi;biometrika50,durbi;biometrika51,durbi;biometrika71}, and the 0.1\% significance $Q$ value~\cite{theil;jasa61}.  These results are presented in Table~\ref{tab;vcrefine}. Serial correlation is more significant when the fit is carried out on a fine grid as $d$ is much smaller than $Q$, but not significant when using a coarse gird. We also note that the non-splitting integration method gives less serial correlation than the splitting method, as its $d$ value is larger, which strengthens the point that the non-splitting method generates less correlated data, although serial correlation is only an indirect measurement of the correlation.

Given that most data analysis software programs, such as Rietveld refinement programs GSAS~\cite{larso;unpub04}, GSAS-II~\cite{toby;jac13} and FullProf~\cite{rodri;unpub90}, do not consider the full VC matrix it is important to minimize error correlations during data analysis since refinements on data that have been processed in a way to minimize the off-diagonal terms in the VC matrix will be the most accurate. In the future, refinement programs that can handle the full VC matrix, coupled with integration protocols that propagate the full VC matrix, may circumvent this issue, at the cost of increased computation time.

We should point out that although the deviation is smaller or comparable with $\Delta_2$, $\Delta_2$ does not give correct estimate for the statistical uncertainty, as is the case for the peak position, where the actual value fluctuations are much smaller. Due to the limitation of the least-square refinement method, it is hard to determine which method gives the correct refined values and uncertainty estimation. Further study, for example, using Bayesian methods may be required to obtain more reliable refinement results.

\section{Conclusions}

This paper discusses methods to extract reliable statistical uncertainties on points in a 1D powder diffraction pattern obtained from widely used 2D integrating detectors. It also explores the origin and extent of statistical correlations between points in the 1D diffraction pattern. The error correlations may be handled correctly by propagating the full VC matrix through the data analysis steps. A software program, {\sc SrXplanar}, is presented for azimuthally integrating 2D detector images while determining statistical uncertainties and the full VC matrix, and for propagating this to the final pattern.  However, most modeling and fitting programs cannot utilize the information in the full VC matrix and so data processing steps that minimize error correlations are explored and an optimal protocol to minimize these correlations is presented. It is strongly suggested to use a non-pixel-splitting integration algorithm and to integrate data directly onto the final 1D grid that will be modeled or further processed. Although the effects of systematic error are larger than the statistical errors in the cases we considered, the true uncertainty may not be determined by systematic errors depending on model deficiencies. Using correct uncertainty information in refinements is important for obtaining correct uncertainty estimation. Failure to do so, for example by neglecting the off-diagonal terms of the VC matrix or fully ignoring the uncertainty information, may result in an underestimation of uncertainties on refined parameters. Estimating uncertainties using $\Delta_2$ defined in Eq.~\ref{eq;covWithoutVC} can account for some contributions of model errors to the uncertainty, but does not give accurate uncertainty estimates in all cases. Even when Eq.~\ref{eq;covWithoutVC} is used to estimate the uncertainties, it is recommended that the correct statistical weights are used in the least-squares equations during model minimization.

\appendix{Supplementary Materials}

\section{A Generalized matrix expression for 2D integration and full VC matrix propagation}
\label{app;vcmatrixPropagation}

Here we generalize the integration equations and other data reduction processes in matrix form. If $\mathbf{p}$ is the vector containing the input information and $\mathbf{T}$ is the transformation matrix that represents a data reduction step, the output information $\mathbf{q}$, in vector form is given by
\begin{equation}
\mathbf{q} = \mathbf{T}\mathbf{p}.
\end{equation}
Assuming the VC matrix of input information is $\mathbf{Cov}_{\mathbf{p}}$, the VC matrix of output information, $\mathbf{Cov}_{\mathbf{q}}$, is given by
\begin{equation}
\label{eq;vcmatrixpropagation}
\mathbf{Cov}_{\mathbf{q}} = \mathbf{T} \ \mathbf{Cov}_{\mathbf{p}} \ \mathbf{T}^{T}.
\end{equation}
The variance on each point of $\mathbf{q}$ is the main diagonal of VC matrix $\mathbf{Cov}_{\mathbf{q}}$, therefore the standard deviation on each point of $\mathbf{q}$ is given by taking square-root of corresponding variance~\cite{princ;b;mticams04,toby;aca04}.

In the integration process, the input vector, transformation matrix and output vector contain the raw counts of each pixel, integration algorithm, and integrated intensities, respectively. In detail, we 'flattened' the 2D raw counts array into a 1D vector using an appropriate flattening method. For example, if the 2D raw counts array has $x\_dimension$ pixels in the $x$ direction and $y\_dimension$ pixels in $y$ direction, the raw counts of pixel ($x$,$y$) may be assigned to the $y \times x\_dimension + x$th element of vector $\mathbf{p}$.

The form of matrix $\mathbf{T}$ is determined by the integration process that is chosen. For example, for the splitting method, the transformation matrix is,
\begin{equation}
\label{eq;Tij}
T_{ij} = \frac{a_{ij}}{\sum_{m} a_{im}},
\end{equation}
where $a_{ij}$ is proportional to the overlap area of the $j$th pixel and $i$th bin. If the $j$th pixel and $i$th bin do not overlap, $T_{ij}$ will equal to zero. For the non-splitting integration method, the equation of the transformation matrix element, $T_{i,j}$ is the same as Eq.~\ref{eq;Tij} , however the $a_{ij}$s can only take the values of 1 or zero depending on whether the pixel center lies in or out of the bin's $2\theta$ range.

The VC matrix of raw counts could be determined using method described in Section~\ref{sec;pixelSwitching} or Section~\ref{sec;esdSingleFrame}. Since we assume the uncertainties estimated in these two methods are statistically independent, the VC matrix of raw counts should be diagonal with the variance equal to the square of the uncertainties of the raw counts.

For re-sampling process, the input vector, transformation matrix and output vector should represent the intensity on the original grid, the re-sampling algorithm, and the intensity on the new grid. Generally, most re-sampling processes can be expressed as linear transformations that transform the diffraction intensities from one base to another base. For example, the simplest two point linear interpolation is
\begin{equation}
T_{m,n} = \frac{x_{p_{n+1}}-x_{q_{m}}}{x_{p_{n+1}}-x_{p_{n}}},
\end{equation}
and
\begin{equation}
T_{m,n+1} = \frac{x_{q_{m}}-x_{p_{n}}}{x_{p_{n+1}}-x_{p_{n}}},
\end{equation}
where $ x_{p_{n}} < x_{q_{m}} < x_{p_{n+1}} $. Other entries of $T$ are zero. Other re-sampling methods may have different elements in the transformation matrix.

\section{Least-squares refinement with full VC matrix}
\label{app;refinevcmatrix}

The least-squares method is usually used in data fitting~\cite{princ;b;mticams04,hahn;b;itc05}. Suppose we have $n$ independent measured data values $y_{i}$, which are believed to be the functions of $m$ variables $p_{j}$. If $m$ is smaller than $n$, we can determine $\mathbf{p}$ by minimizing
\begin{equation}
S = (\mathbf{y} - \mathbf{f}(\mathbf{p}))^T \mathbf{W} (\mathbf{y} - \mathbf{f}(\mathbf{p})),
\end{equation}
where $\mathbf{f}$ is set of functions in our models and $\mathbf{W}$ is the weight matrix. For uncorrelated observed values, $\mathbf{W}$ is a diagonal matrix with $W_{ii}= \sigma_{i}^{-2}$ and for correlated observed values, $\mathbf{W}$ is the inverse of the VC matrix. Given a non-linear model, the best fitting results $\mathbf{p}$ can be determined by iteratively solving the equation
\begin{equation}
\mathbf{A}^{T}\mathbf{W}\mathbf{A} (\mathbf{p} - \mathbf{p}_0) = \mathbf{A}^{T} \mathbf{W} (\mathbf{y} - \mathbf{f}(\mathbf{p})),
\end{equation}
where $\mathbf{p}_{0}$ are the results from the previous iteration. $\mathbf{A}$ is the Jacobian matrix of the model $f$, i.e., $A_{ij} = \frac{\partial f_{i}}{\partial p_{j}}$. The VC matrix of refined parameters $\mathbf{p}$ is the inverse of the Hessian matrix
\begin{equation}
\label{eq;covWithVC}
\mathbf{Cov}_{\mathbf{p}} = (\mathbf{A}^{T}\mathbf{W}\mathbf{A})^{-1}
\end{equation}
and the uncertainty of refined parameters is given by taking the square root of the diagonal values in the VC matrix,  $\sigma[p_i]= \sqrt{(\mathbf{A}^{T}\mathbf{W}\mathbf{A})^{-1}_{ii}}$.

If the VC matrix of data points is unknown, we can do the refinement using the same procedure but with a unity matrix as the weight matrix $\mathbf{W}$. Then the VC matrix of refined parameters can be estimated as
\begin{equation}
\label{eq;covWithoutVC}
\mathbf{Cov}_{\mathbf{p}} = \frac{S}{N-M} (\mathbf{A}^{T} \mathbf{A})^{-1}
\end{equation}
where $N$ and $M$ are number of data points and refined parameters, respectively. However, an implicit assumption here is that the estimation is unbiased only when the model is correct. If the model is incorrect or incomplete, there is no guarantee that uncertainty estimation is unbiased and may either over- or underestimate the uncertainty.



\ack{Acknowledgements}

We would like to thank E. Bo\v{z}in for assistance with the experimental setup and data collection. This work is supported as part of the Center for Re-Defining Photovoltaic Efficiency Through Molecule Scale Control, an Energy Frontier Research Center funded by the U.S. Department of Energy, Office of Science, Office of Basic Energy Sciences under Award Number DE-SC0001085. The APS at Argonne National Laboratory is supported by the U.S. DOE, Office of Science, Office of Basic Energy Sciences, under Contract No. W-31-109-Eng-38.





\bibliography{iucr}
\bibliographystyle{iucr}

\end{document}